\documentclass[aps,prl,nofootinbib,twocolumn,superscriptaddress,letterpaper,preprintnumbers]{revtex4-1}

\usepackage{amssymb}
\usepackage{amsmath}
\usepackage{epsfig}
\usepackage[colorlinks=true
,urlcolor=blue
,anchorcolor=blue
,citecolor=blue
,filecolor=blue
,linkcolor=red
,menucolor=blue
,hyperfootnotes=false,
,linktocpage=true
,pdfproducer=medialab
,pdfa=true
]{hyperref}
\usepackage{array}
\usepackage{booktabs}
\usepackage{comment}
\usepackage{cleveref}
\usepackage{multirow}

\newcolumntype{L}[1]{>{\raggedright\let\newline\\\arraybackslash\hspace{0pt}}m{#1}}
\newcolumntype{C}[1]{>{\centering\let\newline\\\arraybackslash\hspace{0pt}}m{#1}}
\newcolumntype{R}[1]{>{\raggedleft\let\newline\\\arraybackslash\hspace{0pt}}m{#1}}

\usepackage{afterpage}
\usepackage{longtable}
\usepackage{capt-of}

\usepackage{siunitx}
\DeclareSIUnit\electronvolt{e\kern-.05em V}
\DeclareSIUnit\parsec{pc}
\DeclareSIUnit\erg{erg}

\usepackage{graphicx}
\usepackage{gensymb}
\usepackage{subfigure}
\usepackage{blindtext}

\usepackage{listings}
\usepackage[dvipsnames]{xcolor}
\usepackage[T1]{fontenc}

\usepackage{mathtools}
\newcounter{sec}

\newcommand\beq{\begin{alignat}{1}}
\newcommand\eeq{\end{alignat}}
\newcommand{\dhis}{\texttt{DarkHistory} }

\usepackage[normalem]{ulem}

\usepackage{fontawesome}
\newcommand{\nbicon}{{\color{linkcolor}\faFileCodeO}\xspace}
\newcommand{\nblink}[1]{\href{https://github.com/hongwanliu/DarkHistory/tree/T_igm_temperature_constraints/examples/#1.ipynb}{\nbicon}}
\newcommand{\nblinkcc}[1]{\href{https://github.com/hongwanliu/DarkHistory/tree/T_igm_temperature_constraints/examples/Lya_CrossChecks/#1.ipynb}{\nbicon}}
\newcommand{\githubmaster}{\href{https://github.com/hongwanliu/DarkHistory/tree/T_igm_temperature_constraints}{\faGithub}\xspace}

\definecolor{deepgreen}{rgb}{0.2,0.8,0.2}

\colorlet{linkcolor}{BrickRed}

\begin{document}

\preprint{MIT-CTP/5221}

\title{Lyman-$\alpha$ Constraints on Cosmic Heating from Dark Matter Annihilation and Decay}

\author{Hongwan Liu}
\email{hongwanl@princeton.edu}
\affiliation{Center for Cosmology and Particle Physics, Department of Physics, New York University, New York, NY 10003, U.S.A.}
\affiliation{Department of Physics, Princeton University, Princeton, New Jersey, 08544, U.S.A.}

\author{Wenzer Qin}
\email{wenzerq@mit.edu}
\affiliation{Center for Theoretical Physics, Massachusetts Institute of Technology, Cambridge, MA 02139, U.S.A.}

\author{Gregory W. Ridgway}
\email{gridgway@mit.edu}
\affiliation{Center for Theoretical Physics, Massachusetts Institute of Technology, Cambridge, MA 02139, U.S.A.}

\author{Tracy R. Slatyer}
\email{tslatyer@mit.edu}
\affiliation{Center for Theoretical Physics, Massachusetts Institute of Technology, Cambridge, MA 02139, U.S.A.}

\begin{abstract} 
	We derive new constraints on models of decaying and annihilating dark matter (DM) 
	by requiring that the energy injected into the intergalactic medium (IGM) not overheat it
	at late times, when measurements of the Lyman-$\alpha$ forest constrain the IGM temperature.
	We improve upon previous analyses by using the recently developed \texttt{DarkHistory}
	code package, which self-consistently takes into account additional photoionization 
	and photoheating processes due to reionization and DM sources. 
	Our constraints are robust to the uncertainties of reionization and
	competitive with leading limits on sub-GeV DM that decays preferentially to electrons. \githubmaster
\end{abstract}

\maketitle

\noindent
\stepcounter{sec}
\textbf{Introduction.---}Dark matter (DM) interactions such as annihilation or decay can inject a significant amount of energy into the early Universe, producing observable changes in both its ionization and temperature histories.
Changes in the free electron fraction, 
for example, can alter the cosmic microwave background (CMB) anisotropy power spectrum \cite{Adams:1998nr, Chen:2003gz, Padmanabhan:2005es}, allowing constraints on the annihilation cross section \cite{Galli:2009zc,Slatyer:2009yq,Kanzaki:2009hf,Hisano:2011dc,Hutsi:2011vx,Galli:2011rz,Finkbeiner:2011dx,Slatyer:2012yq,Galli:2013dna,Madhavacheril:2013cna,Slatyer:2015jla,Tracy2015_partII}
and the decay lifetime of DM~\cite{Zhang:2007zzh,Slatyer:2016qyl,Poulin:2016anj, Acharya:2019uba} to be set using Planck data~\cite{Planck2018}. 
Constraints based on modifications to the temperature history focus on two redshift ranges where measurement data is or will potentially be available: \textit{(i)} before hydrogen reionization at $z \sim 20$, and \textit{(ii)} during the reionization epoch at $2 \lesssim z \lesssim 6$. In the former redshift range, the 21-cm global signal~\cite{Poulin:2016anj,DAmico:2018sxd,Liu:2018uzy,Cheung:2018vww,Mitridate:2018iag,Clark:2018ghm} and power spectrum~\cite{Evoli:2014pva,Lopez-Honorez:2016sur} have been shown to be powerful probes of DM energy injection, and have the potential to be the leading constraint on the decay lifetime of sub-GeV DM~\cite{Liu:2018uzy}. In the latter range, measurements of the intergalactic medium (IGM) temperature derived from Lyman-$\alpha$ flux power spectra~\cite{Schaye:1999vr, Becker:2010cu} and Lyman-$\alpha$ absorption features in quasar spectra~\cite{Bolton:2011ck, Bolton:2010gr} have been used to constrain the $s$-wave annihilation cross section~\cite{Cirelli:2009bb}, the $p$-wave annihilation cross section, and the decay lifetime of DM~\cite{Diamanti:2013bia, Hongwan2016, Poulin:2016anj}. The IGM temperature can also be used to set limits on the kinetic mixing parameter for ultralight dark photon DM~\cite{Witte:2020rvb, Caputo:2020bdy, Caputo:2020rnx}, the strength of DM-baryon interactions~\cite{Munoz:2017qpy}, and the mass of primordial black hole DM~\cite{PBH2020}. 

In this Letter, we revisit the constraints on $p$-wave annihilating and decaying dark matter from the IGM temperature measurements during reionization.
This work is timely for two reasons. 
First and foremost, the development of \texttt{DarkHistory}~\cite{DarkHistory} allows us to improve on the results of Refs.~\cite{Cirelli:2009bb,Diamanti:2013bia,Hongwan2016} considerably.
We can now self-consistently take into account the positive feedback that increased ionization levels have on the IGM heating efficiency of DM energy injection processes.
This effect can give rise to large corrections in the predicted IGM temperature~\cite{DarkHistory} during reionization. 
Furthermore, \texttt{DarkHistory} can solve for the temperature evolution of the IGM in the presence of both astrophysical reionization sources and dark matter energy injection; previous work only set constraints assuming no reionization~\cite{Cirelli:2009bb}
or a rudimentary treatment of reionization and the energy deposition efficiency~\cite{Diamanti:2013bia, Hongwan2016}.
Second, experimental results published since Refs.~\cite{Cirelli:2009bb,Diamanti:2013bia,Hongwan2016} have considerably improved our knowledge of the Universe during and after reionization. These include: 
\begin{enumerate}
    \item \textit{Planck constraints on reionization}. 
    The low multipole moments of the Planck power spectrum provide information on the process of reionization~\cite{Planck2018}. 
    In particular, Planck provides 68th and 95th percentiles for the ionization fraction in the range $6 \lesssim z \lesssim 30$ using three different models~\cite{Hu:2003gh,Millea:2018bko}, arriving at qualitatively similar results.
    \item \textit{New determinations of the IGM temperature}. By comparing mock Lyman-$\alpha$ power spectra produced by a large grid of hydrodynamical simulations to power spectra calculated~\cite{Walther:2017cir} based on quasar spectra measured by BOSS~\cite{PD:2013gaa}, HIRES~\cite{OMeara:2015lwd,2017AJ....154..114O}, MIKE~\cite{Viel:2013apy}, and XQ-100~\cite{Irsic:2017sop}, Ref.~\cite{Walther:2018pnn} (hereafter Walther\texttt{+}) determined the IGM temperature at mean density in the range $1.8 < z < 5.4$, overcoming a degeneracy between gas density and deduced temperature that hampered previous analyses \cite{Becker:2010cu,Boera:2014sia}. More recently, Ref.~\cite{Gaikwad:2020art} (hereafter Gaikwad\texttt{+}) fit the observed width distribution of the Ly$\alpha$ transmission spikes to simulation results, enabling a determination of the IGM temperature at mean density in the $5.4 < z < 5.8$ redshift range, again with only a weak dependence on the temperature-density relation.
\end{enumerate}

These improvements to both the understanding of energy deposition and the ionization/temperature histories are combined in our analysis into robust constraints on DM $p$-wave annihilation rates and decay lifetimes.
These constraints are competitive in the light DM mass regime ($\lesssim \SI{10}{\giga\eV}$) with existing limits on DM decay from the CMB anisotropy power spectrum~\cite{Slatyer:2016qyl} and are complementary to indirect detection limits~\cite{Essig:2013goa, Cohen:2016uyg, Boudaud:2016mos, Boudaud:2018oya}, being less sensitive to systematics associated with the galactic halo profile and interstellar cosmic ray propagation.

In the rest of this Letter,
we introduce the IGM ionization and temperature evolution equations,
discuss the data and statistical tests used, and finally present our new constraints.  We also include Supplemental Materials that provide additional details to support our main text. For reproducibility, we include links to the code used to generate our figures, indicated by this icon \nbicon.

\noindent
\stepcounter{sec}
\textbf{Ionization and temperature histories.---} In this section, we write down the equations governing the evolution of the IGM temperature, $T_\text{m}$, and the IGM hydrogen ionization level, $x_\text{HII} \equiv n_\text{HII}/n_\text{H}$, where $n_\text{H}$ is the number density of both neutral and ionized hydrogen. The ionization evolution equation is:
\begin{alignat}{1}
    \dot{x}_\text{HII} = \dot{x}_\text{HII}^\text{atom} + \dot{x}_\text{HII}^\text{DM} + \dot{x}_\text{HII}^\star \,.
    \label{eq:ionization_diff_eq}
\end{alignat}
Here, $\dot{x}_\text{HII}^\text{atom}$ corresponds to atomic processes, i.e.\ recombination~\cite{Seager:1999bc, Seager:1999km, Chluba:2010ca, AliHaimoud:2010dx} and collisional ionization, which depend in a straightforward way on the ionization and temperature of the IGM, while $\dot{x}_\text{HII}^\text{DM}$ is the contribution to ionization from DM energy injection. These terms are discussed in detail in Ref.~\cite{DarkHistory}, and are given in full in the Supplemental Materials, as well as a completely analogous HeII evolution equation.
The remaining term, $\dot{x}_\text{HII}^\star$, corresponds to the contribution to photoionization from astrophysical sources of reionization. This term will inevitably source photoheating, which will be important for the IGM temperature evolution equation (discussed below).
$\dot{x}_\text{HII}^\star$ can in principle be determined given a model of astrophysical sources of reionization, but there are large uncertainties associated with these sources. 
For example, the fraction of ionizing photons that escape into the IGM from their galactic sites of production is highly uncertain, ranging from essentially 0 to 1 depending on the model~\cite{McQuinn:2015icp}.

Instead, we rely on the Planck constraints on the process of reionization to fix the form of $\dot{x}_\text{e}$,
allowing us to fix $\dot{x}_\text{HII}^\star$ while remaining agnostic about astrophysical sources of reionization. Specifically, we begin by choosing a late time ionization history, $x_\text{e}^\text{Pl}(z)$ for $z<30$, within the 95\% confidence region determined using either the ``Tanh'' or ``FlexKnot'' model adopted by Planck~\cite{Planck2018}. We then make the common assumption that during hydrogen reionization HI and HeI have identical ionization fractions due to their similar ionizing potentials, but that helium remains only singly ionized due to HeII's deeper ionization potential~\cite{Onorbe2017}.
These assumptions allow us to set $x_\text{HII}^\text{Pl} = x_\text{e}^\text{Pl} / (1+\chi)$, where $\chi \equiv n_\text{He}/n_\text{H}$ is the primordial ratio of helium atoms to hydrogen atoms. 
Given a choice of $x_\text{e}^\text{Pl}(z)$ we can then rearrange Eq.~\eqref{eq:ionization_diff_eq} to set
\begin{alignat}{1}
    \dot{x}_\text{HII}^\star = \left(\frac{\dot{x}_\text{e}^\text{Pl}}{1 + \chi} - \dot{x}_\text{HII}^\text{atom} - \dot{x}_\text{HII}^\text{DM}\right) \theta(z^\star-z) \,,
    \label{eq:photoionization_rate}
\end{alignat}
where $\theta$ is a step function that enforces $\dot{x}_\text{HII}^\star = 0$ at sufficiently early redshifts when astrophysical reionization sources do not exist yet.
To fix $z^\star$, notice that at early times when $\dot{x}_\text{HII}^\star$ is turned off, ionization due to DM energy injection produces $x_\text{e}(z) \geq x_\text{e}^\text{Pl}(z)$.  Since DM cannot significantly reionize the universe~\cite{Hongwan2016}, there will exist a redshift past which $x_\text{e}(z) < x_\text{e}^\text{Pl}(z)$ if we do not turn on $\dot{x}_\text{HII}^\star$.  We define $z^\star$ to be this cross-over redshift where $x_\text{e}(z^\star) = x_\text{e}^\text{Pl}(z^\star)$. 

Thus, for any given DM model and $x_\text{e}^\text{Pl}$ we can use Eq.~\eqref{eq:photoionization_rate} to construct ionization histories that self-consistently include the effects of DM energy injection and reionization simultaneously.
We do not require the astrophysics that produces $\dot{x}_\text{HII}^\star$ to obey any constraint other than $\dot{x}_\text{HII}^\star \geq 0$, which maximizes freedom in the reionization model and leads to more conservative DM constraints.

The IGM temperature history can similarly be described by a differential equation:
\begin{alignat}{1}
    \dot{T}_\text{m} = \dot{T}_\text{adia} + \dot{T}_\text{C} + \dot{T}_\text{DM} + \dot{T}_\text{atom} + \dot{T}^\star \,,
    \label{eq:temp_diff_eq}
\end{alignat}
where $\dot{T}_\text{adia}$ is the adiabatic cooling term, $\dot{T}_\text{C}$ is the heating/cooling term from Compton scattering with the CMB, $\dot{T}_\text{DM}$ is the heating contribution from DM energy injection, and $\dot{T}_\text{atom}$ comprises all relevant atomic cooling processes.
These terms are also fully described in Ref.~\cite{DarkHistory}, and included in the Supplemental Materials for completeness.
We stress that $\dot{T}_\text{DM}$ is computed, using \texttt{DarkHistory}~\cite{DarkHistory}, as a function of both redshift and ionization fraction $x_\text{e}$, self-consistently taking into account the strong dependence of $\dot{T}_\text{DM}$ on $x_\text{e}$, and strengthening the constraints we derive.

The remaining term, $\dot{T}^\star$, accounts for photoheating that accompanies the process of photoionization, as described in Eq.~\eqref{eq:photoionization_rate}. We adopt two different prescriptions for treating the photoheating rate, which we name `conservative' and `photoheated'. In the `conservative' treatment, we simply set $\dot{T}^\star = 0$. This treatment produces highly robust constraints on DM energy injection 
since the uncertainties of the reionization source modeling do not appear in our calculation. Any non-trivial model would only serve to increase the temperature of the IGM, strengthening our constraints. 

\begin{figure*}[!t]
    \centering
    \includegraphics[width=1.0\textwidth]{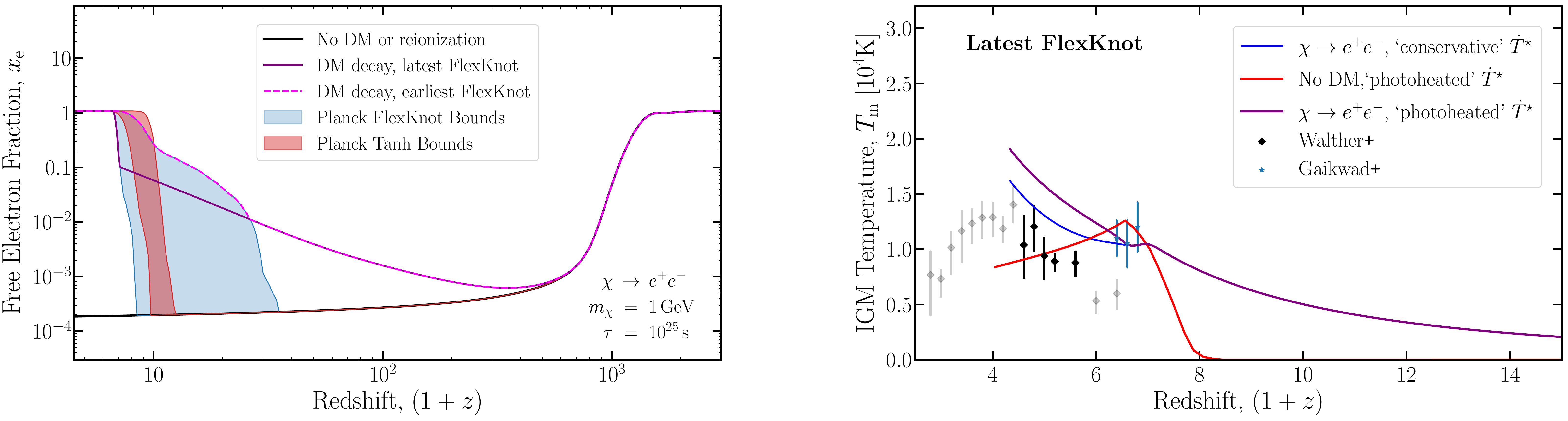}
    \caption{ The ionization history \textit{(Left)} and IGM temperature history \textit{(Right)} as functions of redshift.  The left plot shows the ionization history in the absence of DM energy injection and reionization sources (solid black), the 95\% confidence region for Planck's FlexKnot (shaded blue) and Tanh (shaded red) reionization histories, and the ionization history in the presence of both DM energy injection and reionization sources that produce Planck's latest (solid purple) and earliest (dashed magenta) FlexKnot histories at late times. The right plot shows the temperature history assuming 
\textit{(i)} DM decay and the `conservative' treatment of $\dot{T}^\star$ (blue), \textit{(ii)} the `photoheated' treatment and no DM energy injection (red), and \textit{(iii)} the `photoheated' treatment with DM decay (purple). 
\textit{(i)} and \textit{(iii)} assume a DM mass of $\SI{1}{\GeV}$ and decay to $e^+ e^-$ pairs with a lifetime of $\SI{e25}{\s}$ while 
\textit{(ii)} and \textit{(iii)} assume the latest FlexKnot reionization history and use 
parameter values $(\Delta T, \alpha_\text{bk}) = (\SI{24665}{\kelvin}, 0.57)$ and $(\SI{0}{\kelvin}, 1.5)$, respectively. Also included are the data from Ref.~\cite{Walther:2018pnn} (black diamonds) and Ref.~\cite{Gaikwad:2020art} (blue stars), where the solid data constitute our fiducial data set. \nblink{Lya_plotting}
}
    \label{fig:example_history}
\end{figure*}

In the `photoheated' treatment, we implement a two-stage reionization model.  In the first stage --- prior to the completion of HI/HeI reionization --- we follow a simple 
parametrization adopted in e.g. Refs.~\cite{McQuinn:2015gda,Onorbe2017, Walther:2018pnn} and take $\dot{T}^\star = \dot{x}_\text{HII}^\star (1+\chi) \Delta T$ for some constant $\Delta T$.
This parameter is expected to be within the range \SIrange[range-phrase=--]{2e4}{3e4}{\kelvin} based on analytic arguments~\cite{Miralda_Escude_1994} and simulations~\cite{McQuinn:2012bq, Sanderbeck:2015bba}. 
We will either restrict $\Delta T \geq 0$ or impose a physical prior of $\Delta T\geq \SI{2e4}{K}$ in what we call our `photoheated-I' or `photoheated-II' constraints, respectively.

In the second stage --- after reionization is complete --- the IGM becomes optically thin. In this regime, reionization-only models find that the IGM is, to a good approximation, in photoionization equilibrium~\cite{Puchwein:2014zsa}. The photoheating rate in this limit is specified completely by the spectral index $\alpha_\text{bk}$ of the average specific intensity $J_\nu$ [with units \SI{}{\eV \per \second \per \hertz \per \steradian \per \centi\meter\squared}] of the ionizing background near the HI ionization threshold, i.e. $J_\nu \propto \nu^{-\alpha_\text{bk}}$~\cite{McQuinn:2015gda, Sanderbeck:2015bba}. By considering a range of reionization source models and using measurements of the column-density distribution of intergalactic hydrogen absorbers, the authors of Ref.~\cite{Sanderbeck:2015bba} bracketed the range of $\alpha_\text{bk}$ to be within $-0.5 < \alpha_\text{bk} < 1.5$, which we will use in our analysis.

In summary, the `photoheated' prescription is
\begin{alignat}{1}
    \dot{T}^\star = \begin{dcases}
        \dot{x}_\text{HII}^\star (1 + \chi) \Delta T \,, & x_\text{HII} < 0.99 \,, \\
        \sum_{i\in\{ \text{H}, \text{He} \}} \frac{E_{i\text{I}} x_i}{3(\gamma_{i\text{I}} - 1 + \alpha_\text{bk})} \alpha_{\text{A},i\text{I}} n_\text{H} \,, & x_\text{HII} \geq 0.99 \,,
    \end{dcases}
    \label{eq:modeled_term}
\end{alignat}
%
where $i$ runs over H and He (thus $x_i=$1, $n_\text{He}/n_\text{H}$), 
and for species $i$, $E_{i\text{I}}$ is the ionization potential, 
$\gamma_{i\text{I}}$ denotes the power-law index for the photoionization cross-section at threshold, 
and $\alpha_{A,i\text{I}}$ is the case-A recombination coefficient~\cite{Sanderbeck:2015bba}.
The `photoheated' model is therefore fully specified by two parameters, $\Delta T$ and $\alpha_\text{bk}$.  
Additionally, once HI/HeI reionization is complete, we set $1-x_\text{e} = \SI{4e-5}{}$, which is approximately its measured value~\cite{Bouwens:2015vha}.
This small fraction of neutral HI and HeI atoms 
dramatically decreases the photoionization rate relative to its pre-reionization value for photons of energy $\SI{13.6}{\eV} < E_\gamma < \SI{54.4}{\eV}$ injected by DM. 
Consequently, there is a non-negligible unabsorbed fraction of photons in each timestep, $\exp\left(- \sum_{i \in \{\text{HI}, \text{HeI}\}}n_i \sigma^\text{ion}_i(E_\gamma) \Delta t\right)$, where $\sigma_i^\text{ion}(E_\gamma)$ is the photoionization cross-section for species $i$ at photon energy $E_\gamma$. We modify \dhis to propagate these photons to the next timestep.

\begin{figure*}[t!]
\begin{tabular}{c}
\includegraphics[width=1.0\textwidth]{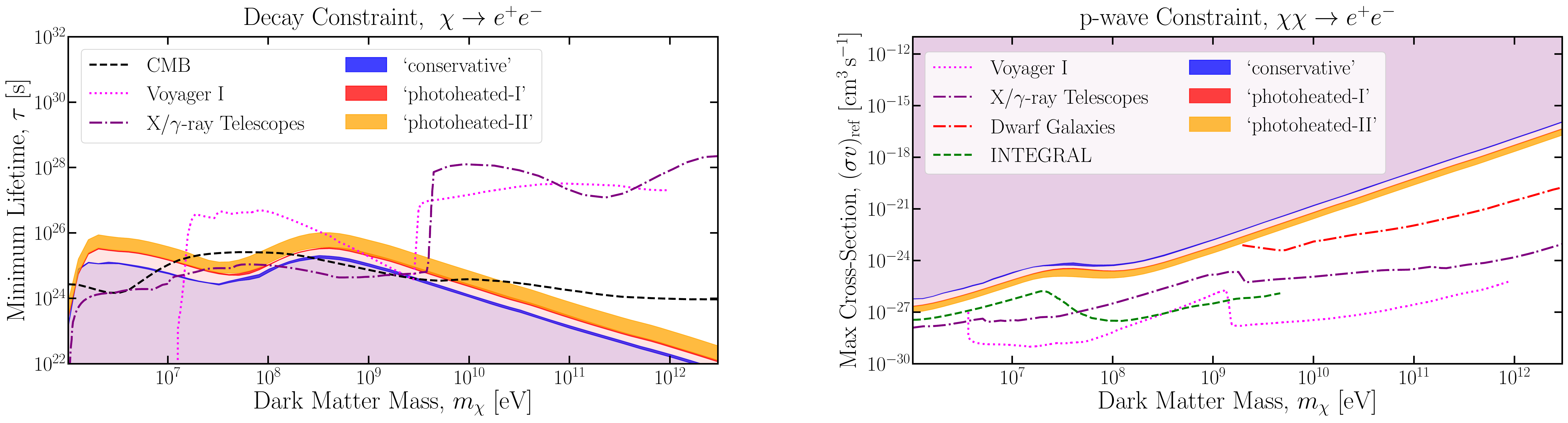}
\end{tabular}
\caption{Constraints for decay (left) or $p$-wave annihilation (right) to $e^+e^-$ pairs with $v_\text{ref} = \SI{100}{\km \, \s^{-1}}$.  We show our constraints using the `conservative' (blue band), `photoheated-I' (red band), or `photoheated-II' (orange band) treatment.
The darkly shaded bands show the variation of our constraints as we vary through the 95\% confidence regions of Planck's Tanh and FlexKnot reionization models.
We also include constraints from the CMB~\cite{Slatyer:2016qyl} (dashed-black), X/$\gamma$-ray telescopes~\cite{Essig:2013goa, Massari:2015xea, Cohen:2016uyg} (dot-dashed purple), INTEGRAL~\cite{Cirelli:2020bpc} where we have assumed $\left< v^2 \right> = \SI{220}{\km \, \s^{-1}}$ in the Milky Way, Voyager I~\cite{Boudaud:2016mos, Boudaud:2018oya} (dotted pink), and
gamma-ray observations of dwarf galaxies~\cite{Zhao:2016xie} (dot-dashed red). \nblink{Lya_plotting}
}
\label{fig:constraints}
\end{figure*}

To demonstrate the effects of DM energy injection and our reionization modeling, we show in Fig.~\ref{fig:example_history} example histories obtained by integrating Eqs.~\eqref{eq:ionization_diff_eq} and~\eqref{eq:temp_diff_eq} for both the `conservative' and `photoheated' treatments, with and without DM decay.  
The left plot shows how our method can produce ionization histories that both take into account the extra ionization caused by DM energy injection and also vary over Planck's 95\% confidence region for the late-time ionization levels.
In the right panel, we assume the Planck FlexKnot curve with the latest reionization, and show in red our best fit temperature history assuming no DM energy injection, with the ‘photoheated’ treatment.
This history is a good fit to the fiducial data, with a total $\chi^2$ of about $5$. 
Additionally, once DM is added we show a model that is just consistent with our (95\% confidence) `conservative' constraints but ruled out by the `photoheated' constraints.

\noindent \textbf{Comparison with data.---} 
\stepcounter{sec}
We compare our computed temperature histories with IGM temperature data obtained from Walther\texttt{+}~\cite{Walther:2018pnn} within the range $1.8 < z < 5.4$ and Gaikwad\texttt{+}~\cite{Gaikwad:2020art} within $5.4 < z < 5.8$.
To construct our fiducial IGM temperature dataset, we only consider data points with redshifts $z > 3.6$ (see Fig.~\ref{fig:example_history}, solid data points) since these redshifts are well separated from the redshift of full HeII reionization \cite{Becker:2010cu},
allowing us to safely use the transfer functions that \dhis currently uses, which assume $x_\text{HeIII}=0$. 
By neglecting HeII reionization and its significant heating of the IGM~\cite{McQuinn:2015icp} we derive more conservative constraints.
Additionally, the two Walther\texttt{+} data points above $z = 4.6$ are in tension with the Gaikwad\texttt{+} result; we discard them in favor of the higher $T_\text{m}$ values reported by Gaikwad\texttt{+}, since this results in less stringent limits. 

To assess the agreement between a computed temperature history and our fiducial temperature dataset using our `conservative' method, we perform a modified $\chi^2$ test. Specifically, our test statistic only penalizes DM models that overheat the IGM relative to the data, which accounts for the fact that any non-trivial photoheating model would only result in less agreement with the data, whereas DM models that underheat the IGM could be brought into agreement with the data given a specific photoheating model.
We define the following test statistic for the $i$th IGM temperature bin: 
\begin{alignat}{1}
    \text{TS}_i = \begin{dcases}
        0 \,, & T_{i,\text{pred}} < T_{i,\text{data}} \,, \\
        \left(\frac{T_{i,\text{pred}} - T_{i,\text{data}}}{\sigma_{i,\text{data}}}\right)^2 \,, & T_{i,\text{pred}} \geq T_{i,\text{data}} \,, 
    \end{dcases}
    \label{eq:one_sided_TS}
\end{alignat}
where $T_{i,\text{data}}$ is the fiducial IGM temperature measurement, $T_{i,\text{pred}}$ is the predicted IGM temperature given a DM model and photoheating prescription, and $\sigma_{i,\text{data}}$ is the $1\sigma$ upper error bar 
from the fiducial IGM temperature data. 
We then construct a global test statistic for all of the bins, simply given by $\text{TS} = \sum_i \text{TS}_i$. Assuming the data points $\{T_{i,\text{data}} \}$ are each independent, Gaussian random variables with standard deviation given by $\sigma_{i,\text{data}}$, the probability density function of $\text{TS}$ given some model $\{T_{i,\text{pred}} \}$ is given by
\begin{multline}
    f(\text{TS}|\{T_{i,\text{pred}}\}) = \frac{1}{2^N} \sum_{n=0}^N \frac{N!}{n! (N-n)!} f_{\chi^2}(\text{TS}; n) \,.
    \label{eq:TS_pdf}
\end{multline}
$N$ is the total number of temperature bins and $f_{\chi^2}(x;n)$ is the $\chi^2$-distribution with argument $x$ and number of degrees-of-freedom $n$, where the $n=0$ case is defined to be a Dirac delta function, $f_{\chi^2}(x;0) \equiv \delta(x)$. The hypothesis that the data $\{T_{i,\text{data}} \}$ is consistent with the $\{T_{i,\text{pred}} \}$ can then be accepted or rejected at the 95\% confidence level based on Eq.~\eqref{eq:TS_pdf}. 
See the Supplemental Materials for more details.

For our `photoheated' constraints, we perform a standard $\chi^2$ goodness-of-fit test.  
For any given DM model we marginalize over the photoheating model parameters by finding the $\Delta T$ and $\alpha_\text{bk}$ values that minimize the total $\chi^2$ subject to the constraints $\Delta T \geq 0$ (`photoheated-I') or $\SI{2e4}{\K}$ (`photoheated-II') and $-0.5 < \alpha_\text{bk} < 1.5$.  We then accept or reject DM models at the 95\% confidence level using a $\chi^2$ test with 6 degrees of freedom (8 data points - 2 model parameters).

Fig.~\ref{fig:constraints} shows constraints for two classes of DM models: DM that decays or $p$-wave annihilates to $e^+e^-$. Our $p$-wave annihilation cross-section is defined by $\sigma v = (\sigma v)_\text{ref} \times (v/v_\text{ref})^2$ with $v_\text{ref} = \SI{100}{\kilo\meter\per\second}$. We also use the NFW boost factor for $p$-wave annihilation calculated in Ref.~\cite{Hongwan2016}.  Although we only show constraints for $e^+e^-$ final states, our method applies to any other final state (see the Supplemental Materials).
The blue, red, and orange regions are excluded by our `conservative,' `photoheated-I,' and `photoheated-II' constraints, respectively. 
The `photoheated' limits are generally a factor of $2-8$ times stronger than the `conservative' constraints. 

The thickness of the darkly shaded bands correspond to the variation in the constraints when we vary $x_\text{e}^\text{Pl}$ in Eq.~\eqref{eq:photoionization_rate} over the 95\% confidence region of Planck's FlexKnot and Tanh late-time ionization curves. 
The `conservative' and `photoheated-I' bands are narrow, demonstrating that the uncertainty in the late-time ionization curve is not an important uncertainty for these treatments. However, the `photoheated-II' treatment shows a larger spread, since the larger values of $\Delta T$ imposed by the prior significantly increase the rate of heating at $z \sim 6$, making the earliest temperature data points more constraining, and increasing the sensitivity to the ionization history at $z \simeq 6$. A better understanding of the process of reionization could therefore enhance our constraints significantly.

Our `conservative' constraints for decay to $e^+ e^-$ are the strongest constraints in the DM mass range $\sim\SI{1}{\MeV} - \SI{10}{\MeV}$ and competitive at around $\SI{1}{\GeV}$ while our $p$-wave constraints are competitive in the range $\sim\SI{1}{\MeV} - \SI{10}{\MeV}$.
For higher masses, constraints from Voyager I observations of interstellar cosmic rays are orders of magnitude stronger for both $p$-wave~\cite{Boudaud:2018oya} and decay~\cite{Boudaud:2016mos}. Constraints from X/$\gamma$-ray telescopes~\cite{Essig:2013goa, Massari:2015xea, Zhao:2016xie, Cohen:2016uyg} are stronger than ours for $m_\chi > \SI{1}{\GeV}$ and comparable for $m_\chi < \SI{1}{\GeV}$.

Importantly, all three types of constraints are affected by different systematics. The telescope constraints are affected by uncertainties in our galactic halo profile while Voyager's are affected by uncertainties in cosmic ray propagation.  
The $p$-wave boost factor is relatively insensitive to many details of structure formation, since it is dominated by the largest DM halos, which are well resolved in simulations (see the Supplemental Materials). 
A more important systematic comes from our assumption of homogeneity. We assume that energy injected into the IGM spreads quickly and is deposited homogeneously, when in reality injected particles may be unable to efficiently escape their sites of production within halos~\cite{Schon:2014xoa, Schon:2017bvu}.
We leave a detailed exploration of these inhomogeneity effects for future work.

\stepcounter{sec}
\noindent \textbf{Conclusion---} 
We have described a method to self-consistently construct ionization and IGM temperature histories in the presence of reionization sources and DM energy injection by utilizing Planck's measurement of the late-time ionization level of the IGM.  We construct two types of constraints for models of DM decay and $p$-wave annihilation.  For the first `conservative' type of constraint, we assume that reionization sources can ionize the IGM but not heat it, resulting in constraints that are robust to the uncertainties of reionization. 
For the second `photoheated' type of constraint, we use a simple but well-motivated photoheating model that gives stronger limits than the `conservative' constraints by roughly a factor of $2-8$.  We expect that as the uncertainties on the IGM temperature measurements shrink, and as reionization and photoheating models become more constrained, these `photoheated' constraints will strengthen considerably. 

\noindent \textbf{Acknowledgments---} 
We thank Sida Lu, Chen Sun and Tomer Volansky for helpful discussions. 
We also thank Gabriele Pezzulli for discussions and for pointing out a typo in v1 of this paper.  
The authors are pleased to acknowledge that the work reported on in this 
paper was substantially performed using the Princeton Research Computing 
resources at Princeton University which is a consortium of groups including the 
Princeton Institute for Computational Science and Engineering and the Princeton 
University Office of Information Technology’s Research Computing department. 
We also thank Siddharth Mishra-Sharma for being the inspiration behind the links 
to GitHub and the Jupyter Notebooks.
HL was supported by the DOE under contract DESC0007968 and the NSF under award PHY-1915409.
WQ was supported by the MIT Department of Physics and a NSF GRFP.
GWR was supported by an NSF GRFP and the U.S. Department of Energy, Office of Science, Office of High Energy Physics and Office of Nuclear Physics under grant Contract Numbers DE-SC0012567 and DE-SC0011090. 
TRS was supported by the U.S. Department of Energy, Office of Science, Office of High Energy Physics, under grant Contract Numbers DE-SC0012567 and DE-SC0013999.

\bibliographystyle{apsrev4-1}
\bibliography{Late_Inj_v2}


\newpage

\onecolumngrid

\begin{center}
\textbf{\large Lyman-$\alpha$ Constraints on Cosmic Heating from Dark Matter Annihilation and Decay} \\ 
\vspace{0.05in}
{ \it \large Supplemental Material}\\ 
\vspace{0.1in}
{}
\vspace{0.05in}
{Hongwan Liu$^{1,2}$, Wenzer Qin$^{3}$, Gregory W. Ridgway$^{3}$, Tracy R. Slatyer$^{3}$}
\end{center}
\centerline{{\it  $^{1}$Center for Cosmology and Particle Physics, Department of Physics, New York University, New York, NY 10003, U.S.A.}}
\centerline{{\it  $^{2}$Department of Physics, Princeton University, Princeton, New Jersey, 08544, U.S.A.}}
\centerline{{\it  $^{3}$Center for Theoretical Physics, Massachusetts Institute of Technology, Cambridge, MA 02139, U.S.A.}}
\vspace{0.2in}

\twocolumngrid

In the following sections 
we provide more detail about the IGM temperature and ionization evolution equations, 
describe several cross-checks that we performed on the results we show in the main text, and 
derive the distribution for the modified $\chi^2$ test used in our `conservative'  constraints. 

\setcounter{sec}{1}
\section{Terms in the Evolution Equations}
\label{app:rates}

In this section we provide explicit expressions for the terms appearing in Eq.~\eqref{eq:ionization_diff_eq} and~\eqref{eq:temp_diff_eq} and explicitly write down the helium ionization evolution equations.  Starting with the non-DM temperature sources,
\begin{alignat}{1}
    \dot{T}_\text{adia} &= -2 H T_\text{m} \,,\nonumber \\ 
    \dot{T}_\text{C} &= -\Gamma_C (T_\text{CMB} - T_\text{m}) \, ,
\end{alignat}
where $H$ is the Hubble parameter, $T_\text{CMB}$ is the temperature of the CMB, and $\Gamma_C$ is the Compton cooling rate 
\begin{align}
    \Gamma_C = \frac{x_\text{e}}{1 +\chi + x_\text{e}} \frac{8 \sigma_T a_r T_\text{CMB}^4}{3 m_\text{e}} \,.
\end{align}
Here, $\sigma_T$ is the Thomson cross section, $a_r$ is the radiation constant, and $m_\text{e}$ is the electron mass.
The DM temperature source is given by
\begin{alignat}{1}
    \dot{T}_\text{DM} &= \frac{2 f_\text{heat}(z, \mathbf{x})}{3(1 + \chi + x_\text{e}) n_\text{H}} \left(\frac{dE}{dV\, dt}\right)^\text{inj}\,
\end{alignat}
where $f_\text{heat}(z, \mathbf{x})$ is the deposition efficiency fraction into heating of the IGM as a function of redshift $z$ and a vector, ${\bf x}$, storing the ionization levels of HI and HeII, which is computed by \texttt{DarkHistory}.  
$\left(\frac{dE}{dV\, dt}\right)^\text{inj}$ is the total amount of energy injected per volume per time through DM decays or annihilations.
Finally, $\dot{T}_\text{atom}$ is given by the sum of the recombination, collisional ionization, collisional excitation, and bremsstrahlung cooling rate fitting functions given in Appendix B4 of Ref.~\cite{Bolton:2006pc}. 
In Fig.~\ref{fig:rates}, we plot these rates for a model of DM decaying to photons with a lifetime of $\SI{2e22}{\s}$ and $m_\chi = \SI{800}{\mega\eV}$.
We set $x_\text{e}^\text{Pl}$ to Planck's latest FlexKnot ionization history and use the `conservative' treatment for the photoheating term. 
Fig.~\ref{fig:rates} demonstrates that in a hot and reionized universe, cooling processes that were once negligible become important and possibly dominant.

\begin{figure}[h!]
\begin{tabular}{c}
\includegraphics[scale=0.48]{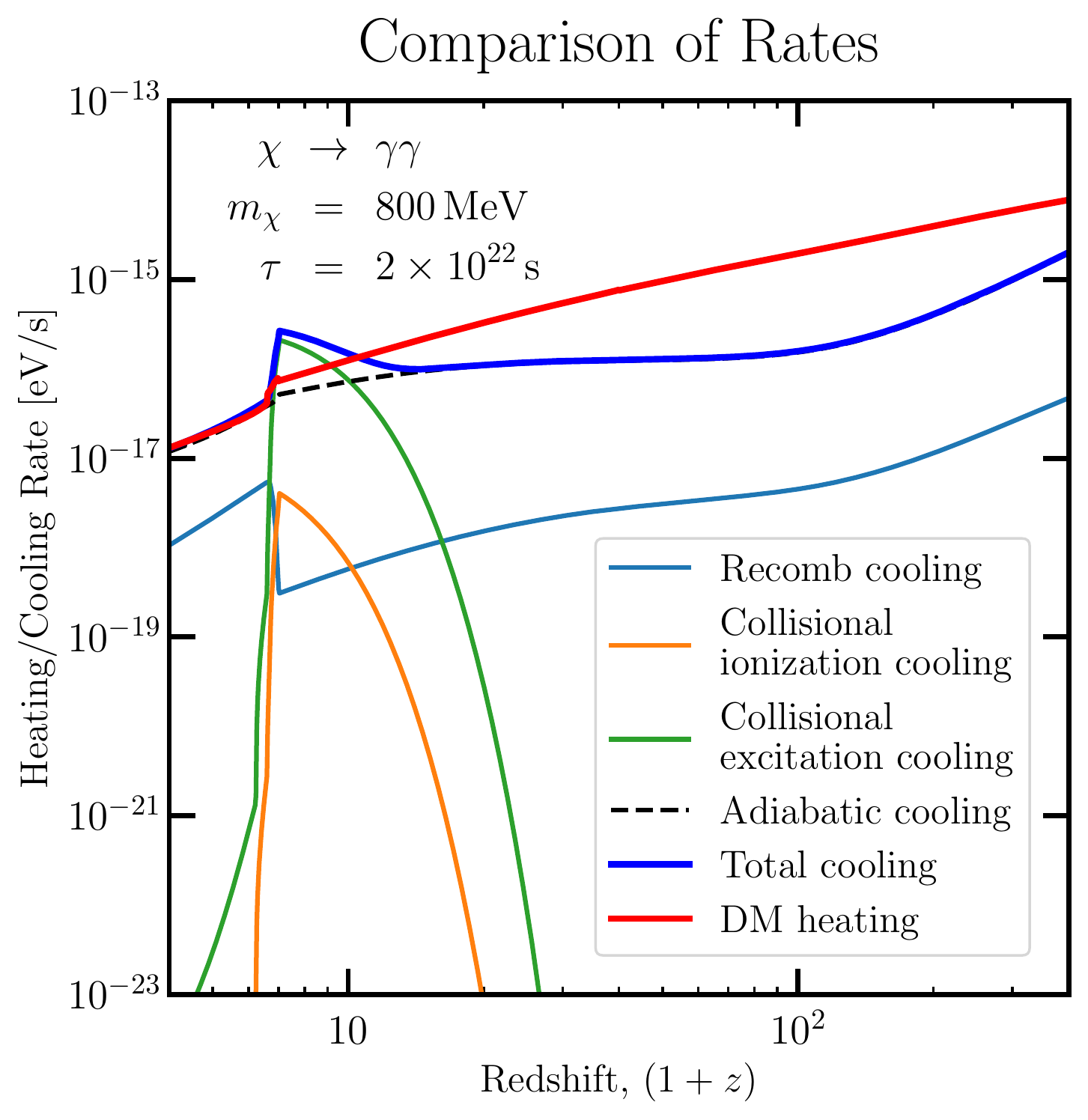}
\end{tabular}
\caption{The absolute value of the atomic cooling rates included in $\dot{T}_\text{atom}$, the adiabatic cooling rate, and the DM heating rate. We assume a model of DM decaying to photons with a lifetime of $\SI{2e22}{\s}$ and $m_\chi = \SI{800}{\MeV}$.  The blue line corresponds to the sum of all cooling rates while the red corresponds to the DM heating rate, the only source of heating in the `conservative' treatment. \nblink{Lya_plotting}
}
\label{fig:rates}
\end{figure}

Moving on to the ionization equations, we write down the helium version of Eq.~\eqref{eq:ionization_diff_eq},
\begin{alignat}{1}
    \dot{x}_\text{HeII} & = \dot{x}_\text{HeII}^\text{atom} + \dot{x}_\text{HeII}^\text{DM} + \dot{x}_\text{HeII}^\star \, , \nonumber \\
    x_\text{HeIII} & = 0 \,,
    \label{eq:He_ionization_diff_eq}
\end{alignat}
where $x_\text{HeII} \equiv n_\text{HeII}/n_\text{H}$ is the density of singly-ionized helium atoms in the IGM normalized to the density of hydrogen atoms, and $x_\text{HeIII}$ is defined similarly. As explained above, the second of these two equations reflects the fact that there are negligibly few fully ionized helium atoms in the IGM over the redshifts under consideration in our analysis. Therefore we only need to keep track of the relative levels of HeI and HeII using the first equation.
Similarly to the $\dot{x}_\text{HII}^\star$ term, we have engineered the astrophysical reionization source term to turn off for $z > z^\star$ and produce a helium ionization curve that is equal to $\frac{\chi}{1+\chi} x_\text{e}^\text{Pl}(z)$ for $z < z^\star$. In other words,
\begin{alignat}{1}
     \begin{dcases}
        \dot{x}_\text{HeII} = \dot{x}_\text{HeII}^\text{atom} + \dot{x}_\text{HeII}^\text{DM} \,, & z > z^\star \,, \\
        x_\text{HeII} = \frac{\chi}{1+\chi} x_\text{e}^\text{Pl}(z) \,, & z < z^\star \,.
    \end{dcases}
    \label{eq:He_photoionization_rate}
\end{alignat}
Notice that we do not need to know the explicit form of $\dot{x}^\star_\text{HeII}$
in contrast to $\dot{x}^\star_\text{HII}$, which we need to compute to evaluate $\dot{T}^\star$ in Eq.~\eqref{eq:modeled_term}.
Due to this simplified treatment, $x_\text{HeII}$ can be discontinuous at $z^\star$; we have tested alternative prescriptions and found negligible effects on our constraints.

The atomic sources contain a contribution from photoionization and a contribution from recombination.
For $z>z^\star$,
we assume a case-B scenario~\cite{Seager:1999bc, Seager:1999km, Wong:2007ym},
\begin{alignat}{3}
	&\dot{x}_\text{HII}^\text{atom} &=& \; 4 \,  \mathcal{C}_\text{H} \, \left[ (1 - x_\text{HII}) \, \beta^B_\text{H} e^{-E_\text{H}/T_\text{CMB}}
	-n_\text{H} \, x_\text{e} \, x_\text{HII} \, \alpha^B_\text{H} \right] \nonumber \\
	&\dot{x}_\text{HeII}^\text{atom} &=& \; 4 \,  \sum_s \mathcal{C}_{\text{HeII}, s} \, \Big[ g_s (\chi - x_\text{HeI}) \, \beta^B_{\text{HeI}, s} e^{-E_{\text{HeI}, s}/T_\text{CMB}} \nonumber \\
	&&& \qquad \qquad \qquad - \, n_\text{H} \, x_\text{e} \, x_\text{HeII} \, \alpha^B_{\text{HeI}, s} \Big] \, ,
\end{alignat}
where $E_i$, $\beta^B_i$, $\alpha^B_i$, and $\mathcal{C}_i$ are, respectively,  the binding energy, case-B photoionization coefficient (including the gaussian fudge factor used in RECFAST v1.5.2~\cite{Seager:1999bc, Seager:1999km}), case-B recombination coefficient, and Peebles $\mathcal{C}_i$ factor for species $i \in \{ \text{H}; \text{HeI, singlet}; \text{HeI, triplet} \}$ \cite{Peebles:1968ja}.
Notice, there is a sum over both spin states of the two electrons in the excited HeI atom. 
For the spin singlet, $g_1 = 1$ and $E_{\text{HeI}, 1} = \SI{20.616}{\eV}$, while for the spin triplet state $g_3 = 3$ and $E_{\text{HeI}, 3} = \SI{19.820}{\eV}$.

When $z < z^\star$, we assume a case-A scenario, which is applicable during reionization~\cite{Bolton:2006pc}:
\begin{alignat}{1}
	\dot{x}_\text{HII}^\text{atom} = & \; n_\text{H}  \, (1-x_\text{HII}) \, x_\text{e}\, \Gamma_\text{eHI} - n_\text{H} \, x_\text{e} \, x_\text{HII} \, \alpha^A_\text{HII} \, .
	\label{eqn:caseA}
\end{alignat}
The collisional ionization rate, $\Gamma_\text{eHI}$, and case-A recombination coefficient, $\alpha^A_\text{HII}$, can be found in Ref.~\cite{Bolton:2006pc}. Notice that the case-A photoionization term from CMB photons is not included because it is exponentially suppressed at these low redshifts, and that the photoionization term from astrophysical reionization sources is already accounted for in $\dot{x}^\star_\text{HII}$.
Additionally, we do not need the analogous HeII version of Eq.~\eqref{eqn:caseA} since at these redshifts we have assumed $x_\text{HeII} = \chi \, x_\text{HII}$.

\begin{figure*}[t!]
\begin{tabular}{c}
\includegraphics[width=0.95\textwidth]{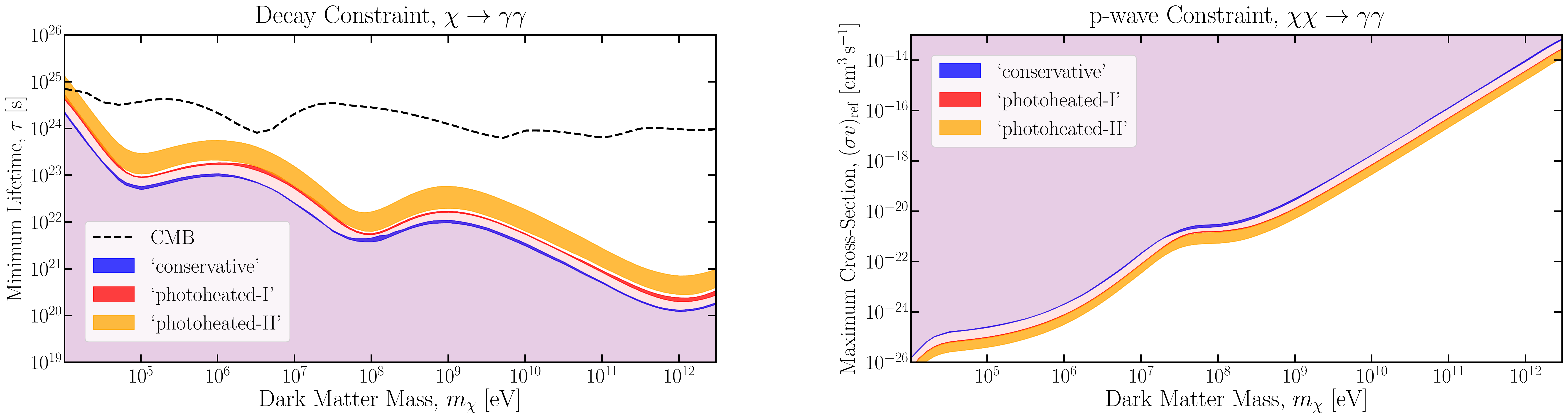}
\end{tabular}
\caption{Constraints for decay (left) or $p$-wave annihilation (right) to $\gamma\gamma$ pairs with $v_\text{ref} = \SI{100}{\km \, \s^{-1}}$.  We show our constraints using the `conservative' (blue band), `photoheated-I' (red band), and `photoheated-II' (orange band) treatments. We also include the CMB constraint for decay~\cite{Slatyer:2016qyl} (dashed-black). Telescope constraints~\cite{Ackermann:2012qk,Boddy:2015efa,Archambault:2017wyh,Abdallah:2018qtu,Acciari:2018sjn} are many orders of magnitude stronger than ours, and are not shown for clarity. \nblink{Lya_plotting}
}
\label{fig:constraints_gg}
\end{figure*}

The DM ionization source terms are given by
\begin{alignat}{1}
	\dot{x}_\text{HII}^\text{DM}   &= \left[\frac{f_\text{H ion}(z, \mathbf{x})}{E_\text{H} n_\text{H}} + \frac{(1 - \mathcal{C_\text{H}}) f_\text{exc} (z, \mathbf{x})}{0.75 E_\text{H} n_\text{H}}\right] \left(\frac{dE}{dV \, dt}\right)^\text{inj} \,, \nonumber \\
	\dot{x}_\text{HeII}^\text{DM} &= \frac{f_\text{He ion}(z, \mathbf{x})}{E_\text{HeI} n_\text{He}} \left(\frac{dE}{dV \, dt}\right)^\text{inj} \,, \nonumber \\ 
	\dot{x}_\text{HeIII}^\text{DM} & = 0 \,,
	\label{eqn:TLA_DM_sources}
\end{alignat}
where $f_\text{H ion}(z, \mathbf{x})$, $f_\text{He ion}(z, \mathbf{x})$, $f_\text{exc}(z, \mathbf{x})$ are the deposition efficiency fractions into hydrogen ionization, single neutral helium ionization, and hydrogen excitation calculated by \texttt{DarkHistory}. 

\section{Other Final States}
\label{app:phot_constr}
In this section we provide constraints for DM decay and $p$-wave annihilation to $\gamma\gamma$, $\mu^+\mu^-$, $\pi^+\pi^-$, and $\pi^0\pi^0$.

\subsection{Photons}
Fig.~\ref{fig:constraints_gg} shows decay and annihilation constraints for $\gamma\gamma$ final states using the `conservative' (blue), `photoheated-I' (red), or `photoheated-II' treatments (orange).
As in the main text, the $p$-wave annihilation cross-section is defined by $\sigma v = (\sigma v)_\text{ref} \times (v/v_\text{ref})^2$ with $v_\text{ref} = \SI{100}{\kilo\meter\per\second}$ and we use the NFW boost factor for $p$-wave annihilation calculated in Ref.~\cite{Hongwan2016}, which accounts for enhanced annihilation due to increased DM density and dispersion velocity in halos. 
Just as in the main text, the darkly shaded blue, red, and orange bands show the variation of our constraints as we vary $x_\text{e}^\text{Pl}$ in Eq.~\eqref{eq:photoionization_rate} over the 95\% confidence region of Planck's FlexKnot and Tanh late-time ionization curves.
As before, the `conservative' and `photoheated-I' bands are narrow, demonstrating an insensitivity to the precise form of the reionization curve, while the `photoheated-II' curve is broader for the reasons discussed in the main text.

The photon final state constraints are less competitive with existing constraints than are the $e^+e^-$ constraints. 
For example, CMB constraints~\cite{Slatyer:2016qyl} are stronger for all masses in the decay channel. 
Additionally, telescope constraints (see e.g. Refs.~\cite{Ackermann:2012qk,Boddy:2015efa,Archambault:2017wyh,Abdallah:2018qtu,Acciari:2018sjn}) 
are many orders of magnitude stronger than ours since telescopes can search directly for the produced photons, in contrast to our temperature constraints that indirectly look for the effects that these photons have on the IGM.

Our $\gamma\gamma$ constraints are weaker than our $e^+e^-$ constraints because the photoionization probability is small (equivalently, the path length is long) for the redshifts and photon energies of interest. In contrast, electrons can efficiently heat the gas either through direct Coulomb interactions (for non-relativistic and mildly relativistic electrons) or through inverse Compton scattering that produces efficiently-ionizing photons (for higher-energy electrons).

\subsection{Muons and Pions}
While we could also consider any other Standard Model particle final state, 
our results from the previous section and the main text indicate that
our constraints are most competitive at masses below $\SI{10}{GeV}$. 
Therefore, we consider 
some of the most important final states that are available to sub-GeV DM besides those already considered: muons, charged pions, and neutral pions. 
To compute the final spectra of $e^+e^-$ and $\gamma$ produced by the decay of pions or muons, we use the PPPC4DMID for DM masses above $\SI{10}{\GeV}$.

For DM masses below $\SI{10}{\GeV}$, we follow the method described in Ref.~\cite{Cirelli:2020bpc}. We start with the spectrum of electrons in the muon rest frame, which is given by
\begin{equation}
	\frac{d N^{\mu \rightarrow e\nu\bar{\nu}}_e}{d E_e} = \frac{4 \sqrt{\xi^2 - 4 \varrho^2}}{m_\mu} [ \xi (3 - 2 \xi) + \varrho^2 (3 \xi - 4)]
\end{equation}
between energies of $m_e < E_e < (m_\mu^2 + m_e^2) / (2 m_\mu)$ and is otherwise zero.
In this equation, $\xi = 2 E_e/m_\mu$ and $\varrho = m_e/m_\mu$.
For a particle $A$ with mass $m_A$ decaying with some spectrum $dN/dE'$ in its rest frame, the spectrum $dN/dE$ in an arbitrary frame where $A$ has energy $E_A$ is given by
\begin{equation}
	\frac{dN}{dE} = \frac{1}{2 \beta \gamma} \int^{E'_\mathrm{max}}_{E'_\mathrm{min}} \frac{dE'}{p'} \frac{dN}{dE'} ,
\end{equation}
where $\gamma = E_A/m_A$ is the Lorentz factor, $\beta = \sqrt{1 - \gamma^{-2}}$, and $E'_{\mathrm{max}/\mathrm{min}} = \gamma (E \pm \beta p)$.
In the case of decay to muons, we can use this equation to boost from the muon frame to the dark matter frame, where the muon has energy $m_\chi$ for annihilations or $m_\chi/2$ for decays.
For decay to pions, we first boost to the pion rest frame where the muon has energy $(m_\pi^2 + m_\mu^2) / (2 m_\pi)$, and then the dark matter frame where the pion similarly has energy $m_\chi$ for annihilations or $m_\chi/2$ for decays.

\begin{figure*}[t!]
\begin{tabular}{c}
\includegraphics[width=0.95\textwidth]{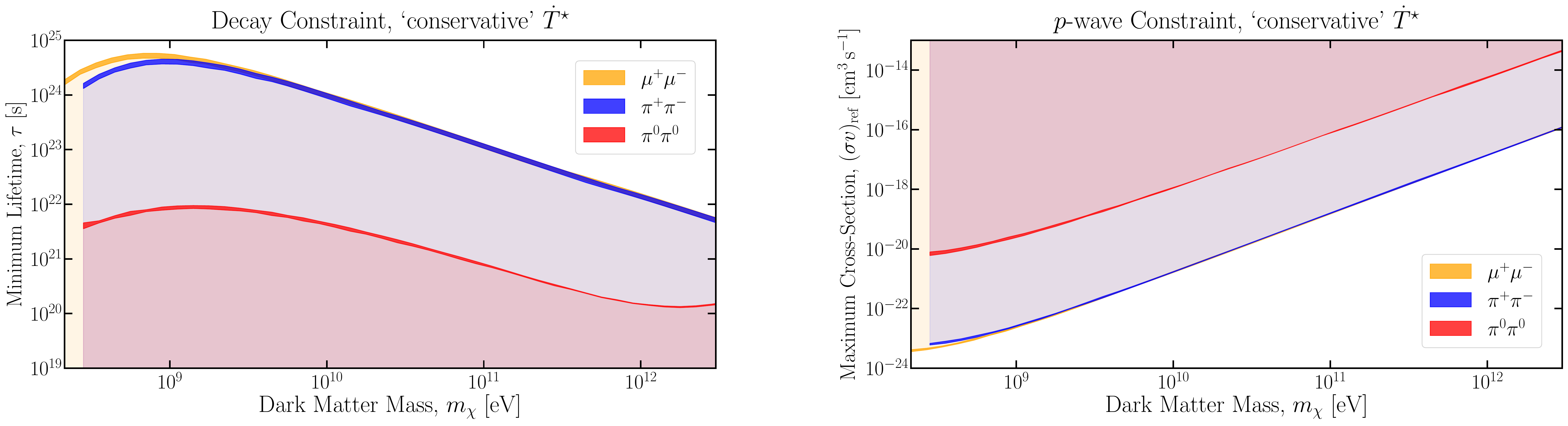}
\end{tabular}
\caption{Constraints for decay (left) or $p$-wave annihilation (right) to $\mu^+\mu^-$ (yellow), $\pi^+\pi^-$ (blue), and $\pi^0\pi^0$ (red) pairs with $v_\text{ref} = \SI{100}{\km \, \s^{-1}}$.  
We show our constraints only using the `conservative' treatments. 
}
\label{fig:constraints_mu_pi}
\end{figure*}

We plot our constraints in Fig.~\ref{fig:constraints_mu_pi}. 
$\mu^+\mu^-$ and $\pi^+\pi^-$ ultimately decay to $e^+e^-$ and neutrinos, meaning that these constraints are comparable to
 the $e^+e^-$ constraints, though somewhat weaker because the produced electrons share at most an $\mathcal{O}(1)$ 
fraction of the total DM injected energy with the other neutrinos. 
$\pi^0\pi^0$ decays almost exclusively to 4$\gamma$, so the photons carry half the energy as compared to photons that result from $\chi \rightarrow \gamma\gamma$ decays. Thus, the pion constraints look exactly like the $\gamma\gamma$ constraints shifted by a factor of 2 to the left.


%
%


\section{Cross Checks}
\label{app:xchecks}
Here, we provide cross checks to validate the assumptions we made in our analysis. 
First, we will validate maintaining $x_\text{HeIII} = 0$ after H and HeI reionization despite DM injecting HeII ionizing photons. 
Second, we will check that our $p$-wave constraints are insensitive to the uncertainty in the halo boost factor coming from the halo profile. Finally, we will validate our use of ionization histories that feature significant ionization levels prior to reionization, by checking that they do not violate constraints on the total $z < 50$ optical depth.

\subsection{Treatment of HeIII}
\label{sec:HeIII}
In calculating the constraints shown in Fig.~\ref{fig:constraints}, we assume that there is no ionization of HeII to HeIII --  i.e. $x_\text{HeIII}=0$ -- consistent with the assumptions that went into the making of \texttt{DarkHistory}'s transfer functions. 
We still account for energy deposition through ionization of HeII by
allowing photons with energies $E_\gamma > \SI{54.4}{\eV}$ to be absorbed by HeII atoms, thus producing electrons of energy $E_\gamma - \SI{54.4}{\eV}$ that thermalize with the IGM.
This is not entirely self-consistent because these photoionization events would gradually increase the fraction of HeII atoms as they convert into HeIII atoms.  
Having fewer HeII atoms could then affect our constraints by decreasing the heating deposition fraction, since fewer photoionized electrons could be produced and thermalize with the IGM.

We test our sensitivity of our constraints to this approximation by adding a new $\dot{x}_\text{HeIII}$ source term and accounting for recombination photons once HI/HeI reionization is complete.
We restrict this correction to after HI/HeI reionization
because it is expected to make the biggest difference in the heating rate then, since HeII atoms are the only possible source of photoionized electrons at this point, and because the temperature data we use are primarily in this redshift range.

To apply our correction,
we first modify Eq.~\eqref{eq:He_ionization_diff_eq} to track the fully ionized helium fraction,
\begin{alignat}{1}
	\dot{x}_\text{HeIII} = & \frac{f_\text{He ion}(z, \mathbf{x})}{4 E_\text{H} n_\text{He}} \left(\frac{dE}{dV \, dt}\right)^\text{inj} \nonumber \\
	& + n_\text{H}  \, (\chi-x_\text{HeIII}) \, x_\text{e}\, \Gamma_\text{eHeII} \nonumber \\
	& - n_\text{H} \, x_\text{e} \, x_\text{HeIII} \, \alpha^A_\text{HeIII} \, ,
\end{alignat}
where the deposition fraction $f_\text{He ion}(z, \mathbf{x})$ computed by \dhis accounts for the total energy deposited into HeII ionization
and $\Gamma_\text{eHeII}$ is the collisional ionization rate of HeII~\cite{Bolton:2006pc}. 
We then compute the fraction of HeIII atoms that recombines within a timestep of the code, $\Delta t$,
\begin{alignat}{1}
	f^\text{HeIII}_\text{recomb} = 1 - e^{-\alpha^{\text{A}}_{\text{HeIII}} x_\text{HeIII} n_e \Delta t} \, .
\end{alignat}
We convert this fraction to the number of $\SI{54.4}{\eV}$ photons per baryon emitted by HeIII atoms in this time step
\begin{alignat}{1}
	N^\text{HeIII}_\text{recomb} = f^\text{HeIII}_\text{recomb} n_\text{HeIII}/n_\text{B} \, ,
\end{alignat}
then add these photons to \texttt{DarkHistory}'s low energy photon spectrum within that time step.

\begin{figure}[h]
	\begin{tabular}{c}
		\includegraphics[scale=0.38]{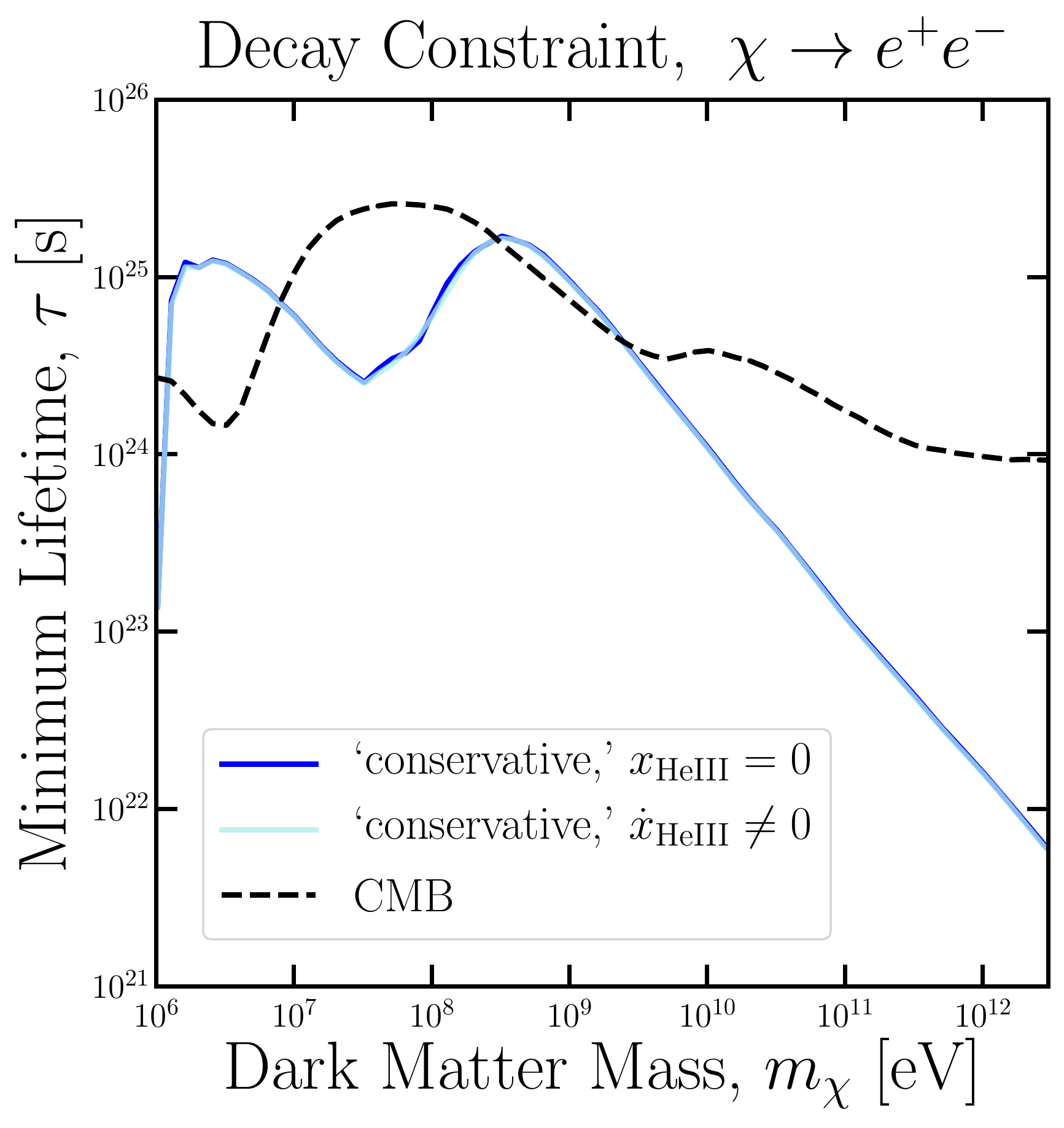}
	\end{tabular}
	\caption{Comparison of `conservative' constraints for decay to electrons, including and not including the effects of HeII ionization after HI/HeI reionization. Both constraints were generated assuming Planck's earliest Tanh reionization history. \nblinkcc{IGM_checks_CleanVersion}
	}
	\label{fig:HeIII}
\end{figure}

Fig.~\ref{fig:HeIII} shows a comparison of constraints for dark matter decaying to electrons, where the two curves either allow for a non-zero HeIII fraction (light blue) or do not (blue). 
The difference in constraints is always less than 1\%, and so is not an important source of error in our analysis.

\subsection{Boost factor for $p$-wave annihilation}
\label{sec:pwave_boost}
The boost factor due to enhanced density and velocity dispersion in halos depends on the halo profile chosen. However, in Ref.~\cite{Hongwan2016}, the boost factor was found to be highly robust to this choice, since the main contribution to the boost factor comes from the largest halos, which are fully resolved in $N$-body simulations.
We find that the difference in our constraints made by using the Einasto $p$-wave boost factor rather than the NFW $p$-wave boost factor from Ref.~\cite{Hongwan2016} is negligibly small, resulting in a modification of no more than 0.5\% to our constraints.
Notice that the two boost factors only vary over the halo mass function and halo profile, and do not include uncertainties due to mergers, asphericity, etc.  

\subsection{Optical depth}
\label{sec:opdepth}

In this section, we discuss the relation between temperature and ionization constraints, focusing in particular on the complementarity of these constraints. One might worry that scenarios excluded by excess heating of the IGM are strictly a subset of those excluded by the ionization history.
In some cases, the DM contribution to the optical depth $\tau$ before reionization, combined with one of the Planck reionization models, can exceed the Planck limit on $\tau$. DM energy injection starts to increase the ionization fraction and temperature immediately after recombination, and so our computed ionization histories will always be in excess of Planck's reionization curves at early enough redshifts.

To some extent, these worries have already been addressed by the fact that the temperature constraints can sometimes be stronger than the CMB power spectrum constraints for DM decays as derived in Refs.~\cite{Slatyer:2016qyl}, which account for the effect of excess ionization on the full multipole structure of the CMB power spectrum. For simplicity, however, we would like to compare the IGM temperature constraints derived in the main body with limits on the ionization history coming simply from the Planck upper limit on $\tau$. 

Given an ionization history $x_\text{e}(z)$, the optical depth is
\begin{equation}
	\tau = n_{\text{H,0}}  \sigma_T \int_0^{z_\mathrm{max}} dz \, x_\text{e}(z) \frac{(1+z)^2}{H(z)},
	\label{eqn:opt_dpth}
\end{equation}
where $\sigma_T$ is the Thomson cross-section and $z_\mathrm{max}$ is set to 50, as is done in Ref.~\cite{Planck2018}. The 68\% upper bound on the optical depth from Planck assuming a tanh function reionization history is $\tau = 0.0549$~\cite{Planck2018}. To derive a constraint, 
we compute an ionization history in the presence of DM energy injection 
and exclude it if the history's optical depth is greater than $0.0549$.

Clearly, these optical depth constraints will be highly sensitive to the reionization curve we choose. 
For example, if we were to use the earliest Tanh reionization curve that already saturates the optical depth bound we would rule out all DM models since they all increase $\tau$.
On the other hand, we saw that our temperature constraints were very weakly dependent on the choice of reionization curve.
For a fair comparison, we choose a reionization history with the smallest optical depth.
While we could choose the latest Tanh reionization curve, we instead follow the instantaneous reionization method described in Ref.~\cite{Hongwan2016} so that we can compare to older optical depth constraints.
We will assume an instantaneous HI/HeI reionization at $z=6$, then an instantaneous HeII reionization at $z=3$, but no other sources of reionization other than DM for $z>6$.
The optical depth contributed by the range $0 < z < 6$ is $0.384$,\footnote{This is nearly equivalent to using the earliest Tanh reionization history, which has an optical depth contribution of $0.383$ to the same redshift range.}
meaning that DM models that contribute more than $\delta \tau = 0.0165$ to the optical depth within the range $6 < z < 50$ will be ruled out.

\begin{figure}[h]
	\begin{tabular}{c}
		\includegraphics[scale=0.48]{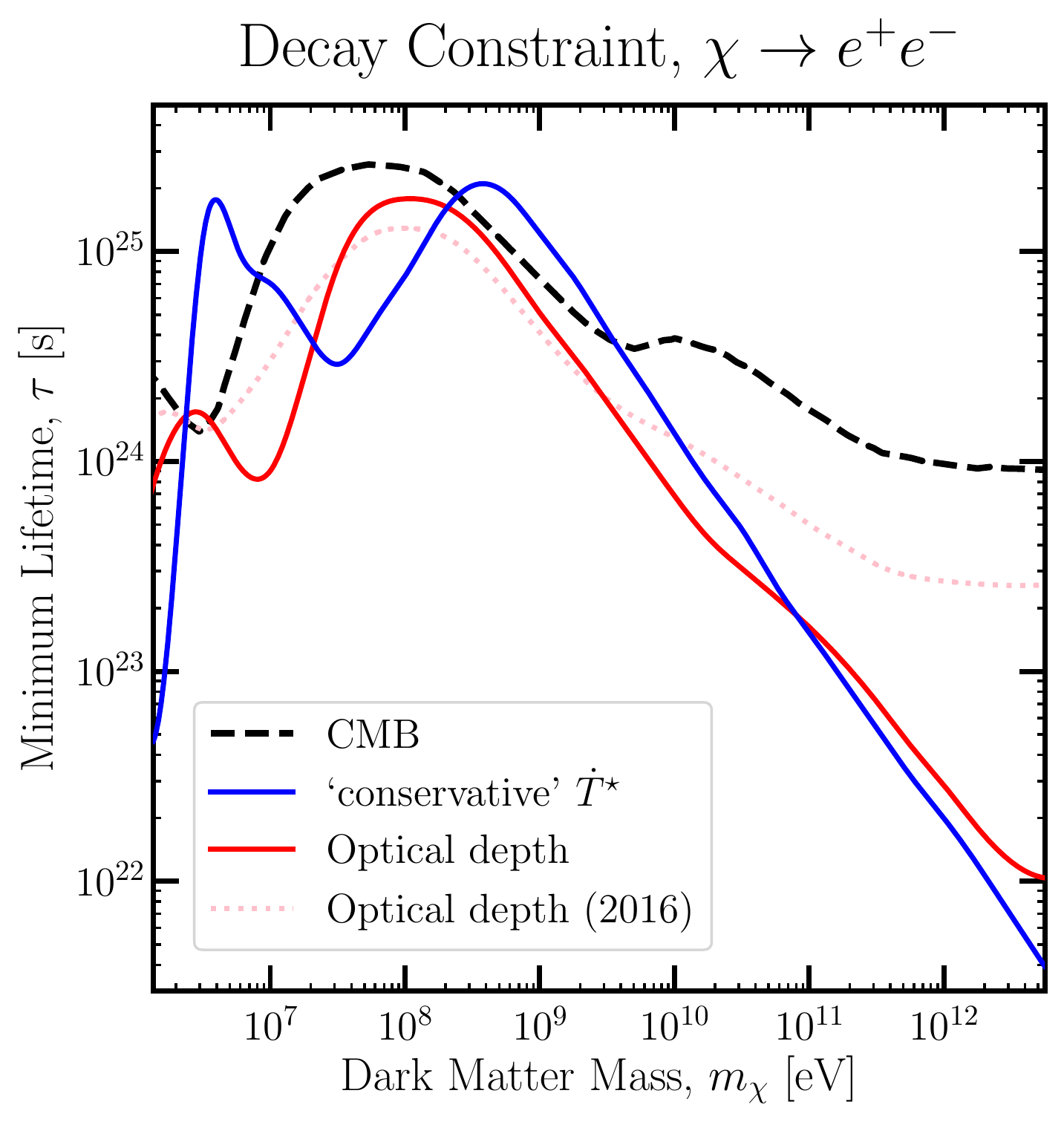}
	\end{tabular}
	\caption{Constraints obtained from the IGM temperature (red) and optical depth (blue, solid), as well as previous bounds derived in Ref.~\cite{Hongwan2016} from the optical depth (pink, dashed). The black line shows the constraints derived using a principal component analysis of CMB data~\cite{Slatyer:2016qyl}. \nblinkcc{IGM_checks_CleanVersion}
}
	\label{fig:opdepth}
\end{figure}

Fig.~\ref{fig:opdepth} shows a comparison of the optical depth constraint in blue to the IGM temperature constraint in red, as well as to a previous constraint made with Planck intermediate results ~\cite{Hongwan2016}, which measured $\tau = 0.058 \pm 0.012$ and is represented by the dashed curve~\cite{Adam:2016hgk}. We see that across most of the mass range, the two methods of constraining dark matter parameters are comparable, but there is a large range of DM masses over which the temperature constraints do better than the optical depth limits. To summarize, since the IGM temperature constraints are insensitive to the exact ionization history during reionization, they probe a different aspect of energy injection from DM that is distinct from ionization-based constraints like optical depth and the CMB power spectrum. Finally, we show for reference an older optical depth constraint across all masses~\cite{Hongwan2016}, which calculated $\delta \tau$ by integrating over the excess ionization fraction over the standard three-level atom result up to recombination, following the method in Ref.~\cite{Cirelli:2009bb}. 

\section{Test Statistics}
\label{app:stats}

In this section, we derive the distribution of the modified $\chi^2$-like test statistic (TS) that we use in conjunction with the `conservative' treatment of the  $\dot{T}^\star$ photoheating term (i.e. $\dot{T}^\star=0$). We are working in a frequentist framework, so we wish to evaluate the probability distribution for the TS defined in Eq.~\eqref{eq:one_sided_TS}, when assuming a certain pattern of heating due to DM energy injection. We can then say that this scenario is excluded if the TS observed in the real data is sufficiently unlikely. We make the assumption that the data points in different redshift bins are independent and Gaussian distributed.

Suppose that there are $N$ redshift bins, and in the $i$th bin the temperature value $T_{i,\text{data}}$ is drawn from a Gaussian distribution with mean $T_{i,\text{pred}}$ and standard deviation $\sigma_{i,\text{data}}$. There is then a $50\%$ chance that $T_{i,\text{data}} > T_{i,\text{pred}}$, so the probability distribution for $\text{TS}_i$ as defined in Eq.~\eqref{eq:one_sided_TS} is:

\begin{align} 
	f(\text{TS}_i | T_{i,\text{pred}})& = \frac12 \delta(\text{TS}_i) + P(T_{i,\text{data}}) \frac{d(T_{i,\text{data}})}{d(\text{TS}_i)}  \nonumber \\
	& = \frac12 \delta(\text{TS}_i) + \frac{1}{2 \sqrt{2 \pi}} \text{TS}_i^{-1/2} \exp(-\text{TS}_i/2) \,,
\end{align}

where $\delta$ is the Dirac delta function. Let the $\chi^2$ probability distribution function with $j$ degrees of freedom be denoted by $f_{\chi^2}(\text{TS}; j)$. Then one can rewrite this distribution in terms of the $\chi^2$ distribution with one degree of freedom:

\begin{align} 
	f(\text{TS}_i|T_{i,\text{pred}}) & = \frac12 \delta(\text{TS}_i) + \frac12 f_{\chi^2}(\text{TS}_i; 1) \, .
\end{align}

Now we want to know the distribution for the total TS value from combining the bins (assuming uncorrelated data), $\text{TS} \equiv \sum_i \text{TS}_i$. We can write:

\begin{alignat}{2}
    f(\text{TS}|\{T_{i,\text{pred}} \}) &=&& \left[ \prod_{i=1}^N \int_0^\infty d\text{TS}_i \,f(\text{TS}_i | T_{i,\text{pred}}) \right] \nonumber \\
    & && \,\, \times  \delta (\text{TS} - \sum_{j=1}^N \text{TS}_j )  \nonumber \\
    &=&& \left[ \prod_{i=1}^N \int_0^\infty d\text{TS}_i \, \left[ \delta(\text{TS}_i) + f_{\chi^2} (\text{TS}_i;1) \right] \right] \nonumber \\
    & && \,\, \times \frac{1}{2^N}  \delta (\text{TS} - \sum_{j=1}^N \text{TS}_j ) \,.
\end{alignat}

Expanding the product inside the integrals gives a sum of terms, which each consist of a product of delta-functions and $f_{\chi^2}$ functions. For a term with $n$ delta-functions, the delta-functions can be used to do $n$ of the integrals, resulting in a term of the form:
\begin{equation} 
	\left[ \prod_{j=n+1}^N \int_0^\infty d \text{TS}_{i_j}  f_{\chi^2}(\text{TS}_{i_{j}}; 1)\right] \delta(\text{TS} - \sum_{k=n+1}^N \text{TS}_{i_k})\,,
\end{equation}
where $n$ can take on values from $0$ to $N$, and $i_{n+1}, i_{n+2}, \cdots, i_N$ are a collection of indices between 1 and $N$ for this particular term. However, this is exactly the standard probability distribution function for the sum of the $\chi^2$ test statistic over $N-n$ bins, so we can write it as $f_{\chi^2}(\text{TS}; N-n)$.

The coefficient of each such term will be the number of ways of choosing which $n$ indices correspond to $\delta$-function terms as opposed to the $N-n$ indices labeling $f_{\chi^2}(\text{TS}_{i}; 1)$ contributions -- which is the binomial coefficient $\binom{N}{n}$. Since $\binom{N}{n}$ = $\binom{N}{N-n}$, we can write:
\begin{equation} 
	f(\text{TS}|\{T_{i,\text{pred}}\})  =  \frac1{2^N} \sum_{n=0}^N \binom{N}{n} f_{\chi^2}(\text{TS}; n) \, .
\end{equation}
This completes the proof of Eq.~\eqref{eq:TS_pdf}.

Note that this expression integrates correctly to 1, as $\int d \text{TS} f_{\chi^2}(\text{TS}; n)= 1$ and $\sum_{n=0}^N \binom{N}{n} = 2^N$.
The largest binomial coefficients $\binom{N}{n}$ will occur for $n\approx N/2$, and so we may approximate the distribution as a $\chi^2$ distribution with $N/2$ degrees of freedom. However, for the actual constraints in the main text we use the full distribution, rather than this approximation.

We can also understand this distribution by thinking of $\text{TS}$ as the standard $\chi^2$ test statistic, in the presence of a model for the data where each redshift bin contains an irreducible (dark matter) contribution plus a non-negative but otherwise arbitrary increase to the temperature from photoheating. If we profile over the nuisance parameters describing the unknown astrophysics, we see that the minimum $\chi^2$ will be attained when:
\begin{itemize}
\item in bins where the irreducible contribution from dark matter already exceeds the measured temperature, extra contributions from photoheating are set to zero; the contribution to the TS is the usual $\chi^2$ computed using the irreducible model and the data,
\item in bins where the irreducible contribution from dark matter does not exceed the measured temperature, the additional photoheating contribution is chosen to precisely match the data, and consequently the contribution to the TS is zero.
\end{itemize}
This is exactly the prescription for our modified TS, Eq.~\eqref{eq:one_sided_TS}. 

Because this is a standard $\chi^2$ test, just with a flexible background model, the probability distribution for the TS should follow that of a $\chi^2$ distribution with $N-m$ degrees of freedom, where $m$ is the number of floated parameters in the fit. The number of floated parameters for this signal model is the number of bins where the data is greater than the irreducible model, which can vary from 0 to $N$; thus the full probability distribution is obtained as a linear combination of $\chi^2$ distributions with degrees of freedom varying from $0$ to $N$.

\end{document}